\documentclass[twocolumn]{el-author}

\newcommand{\ua}{\uparrow}
\newcommand{\nc}{\newcommand}
\nc{\da}{\downarrow} \nc{\hc}{\hat{c}} \nc{\hS}{\hat{S}}
\nc{\bra}{\langle} \nc{\ket}{\rangle} \nc{\eq}{equation (\ref}
\nc{\h}{\hat} \nc{\hT}{\h{T}}\nc{\be}{\begin{eqnarray}}
	\nc{\ee}{\end{eqnarray}}\nc{\rd}{\textrm{d}}\nc{\e}{eqnarray}\nc{\hR}{\hat{R}}\nc{\Tr}{\mathrm{Tr}}
\nc{\tS}{\tilde{S}}\nc{\tr}{\mathrm{tr}}\nc{\8}{\infty}\nc{\lgs}{\bra\ua,\phi|}\nc{\rgs}{|\ua,\phi\ket}
\nc{\hU}{\hat{U}}\nc{\lfs}{\bra\phi|}\nc{\rfs}{|\phi\ket}\nc{\hZ}{\hat{Z}}\nc{\hd}{\hat{d}}\nc{\mD}{\mathcal{D}}
\nc{\bd}{\bar{d}}\nc{\bc}{\bar{c}}\nc{\mc}{\mathcal}\nc{\ea}{eqnarray}\nc{\mG}{\mathcal{G}}\nc{\bce}{\begin{center}}
	\nc{\ece}{\end{center}}
\date{25th May 2021}

\begin{document}
	
	\title{On the Sparsity Bound for the Existence of a Unique Solution  in Compressive Sensing  by the Gershgorin  Theorem}
	
	\author{Ljubiša Stanković}
	
	\abstract{Since compressive sensing deals with a signal reconstruction using a reduced set of measurements, the existence of a unique solution is of crucial importance. The most important approach to this problem is based on the restricted isometry property which is  computationally unfeasible. The coherence index-based uniqueness criteria are computationally efficient, however, they are pessimistic. An approach to alleviate this problem has been recently introduced by relaxing the coherence index condition for the unique signal reconstruction using the orthogonal matching pursuit approach. This approach can be further relaxed and the sparsity bound improved if we consider only the solution existence rather than its reconstruction. One such improved bound for the sparsity limit is derived in this paper using the Gershgorin disk theorem.  
	}
	
	\maketitle

\section{Introduction}

In compressive sensing we are dealing with a reduced set of signal observations \cite{donoho2006,sur,Acces,Wiley,Tutorial,candes2006,OMP_2,MS_1,INg,MS_3,GradientIET}. The reduced set of observations can be caused by a desire to acquire a signal with a low number of observations or by
physical unavailability to measure the signal at all possible sampling positions and to get a complete set of samples \cite{Wiley,Tutorial}. In some applications, signal samples may be heavily
corrupted at some arbitrary positions that their omission could be the best approach to their processing, when we are left with a reduced set of signal samples to reconstruct the signal  \cite{Impulsive,DCT,IMIMNCS}.  The main condition to fully reconstruct the signal from a reduced set of observations is the signal sparsity in a transformation domain. Sparse signals can be reconstructed from reduced measurements under some conditions \cite{candes2006,RIP,RIPP,Uniq}. Applications of 
compressive sensing methods are numerous, including radar signal processing \cite{LJIM,radarCS3}, time-frequency analysis \cite{TFCS2,ISAR22,H_6}, data hiding \cite{TFCS4}, wireless communications \cite{WirC}, and image processing \cite{Gradient_Isi}.

While compressive sensing provides a basis for signal reconstruction, assuming the sparsity in a transformation domain, the uniqueness of the solution is of crucial importance, due to the reduced set of measurements. The most comprehensive uniqueness condition has been defined through the restricted isometry property that is computationally not feasible. An alternative approach is based on the coherence index. However, this criterion may be quite pessimistic. 

An approach to improve the coherence index-based bound has been proposed in \cite{CB} by analyzing the initial estimate and the support uncertainty principle as in \cite{UN,elad-bases}. The approach presented in \cite{CB} guarantees unique reconstruction of a sparse signal using the orthogonal matching pursuit approach. In this paper, a relaxed coherence index condition will be derived for the existence of the unique solution of the compressive sensing problem, using the Gershgorin disk theorem. This result guarantees the unique solution existence, but not its reconstruction, meaning that the obtained bound can be relaxed as compared to the one introduced in \cite{CB}. The new result for the sparsity bound will be related to the classical one and those proposed in \cite{CB}, as well as  illustrated on numerical examples.

\index{Sparse signals}%
\index{Compressive sensing}

\section{Definitions}

Consider an $N$-dimensional signal in one of its transformation domains, with elements denoted as $$\mathbf{X}=[X(0), ~X(1),...,~X(N-1)]^T,$$ where $T$ represents
the transpose operation. The signal is sparse in the considered transformation
domain if the number of nonzero elements, denoted by $K$, is much smaller than the
signal dimension, $N$, that is, if the following property holds
$$X(k)=0 
\text{ \ \ for \ \ }
k\notin \mathbb{K}=\{k_{1},k_{2},...,k_{K}\}$$ 
and $K\ll N$. 
The number of
nonzero elements can be expressed using the $\ell_{0}$-norm operator or the set cardinality operator as 
\[
\left\Vert \mathbf{X}\right\Vert _{0}=\mathrm{card}\left\{  \mathbb{K}
\right\}  =K,
\]
where $\left\Vert \mathbf{X}\right\Vert
_{0} $ is the $\ell_{0}$-norm (norm-zero) and  $\mathrm{card}\left\{  \mathbb{K}\right\}$ is the cardinality of set
$\mathbb{K}.$ 

The observations or measurements of the sparsity domain elements are defined as their linear combinations
\begin{equation}
y(m)=\sum_{k=0}^{N-1}a_{k}(m)X(k), \label{Hip00}%
\end{equation}
where $m=0,1,\ldots, M-1$ is the index of a measurement and $a_{k}(m)$, $k=0,1,\dots,N-1$, are
the weighting coefficients of the $m$-th measurement. 
The measurement vector, $\mathbf{y}$, is given by
\[
\mathbf{y=[}y(0),~y(1),~...,y(M-1)]^{T}.
\]
Within the framework of linear systems of equations, the measurements can be considered as an undetermined system with
$M<N$ equations
\[
\mathbf{y=AX},
\]
where $\mathbf{A}$ is the measurement matrix with elements $a_k(m)$. The  size of the measurement matrix is $M\times N$.

The fact that the signal must be sparse in a transformation domain, with $X(k)=0$ for $k\notin \mathbb{K}=\{k_1,k_2,...,k_K\}$,
is not taken into account within the measurement matrix
$\mathbf{A}$ since, in general, the positions of the nonzero values of $X(k)$ are unknown and should be determined. If we assume that the nonzero positions are found (assumed or known in advance), meaning that $X(k)=0$ for $k\notin \mathbb{K}$, then a  system with a reduced number of unknowns is obtained. This system corresponds to a reduced $M\times K$ measurement matrix $\mathbf{A}_{K}$. The system of equations then assumes the form
\begin{equation}
\mathbf{y}=\mathbf{A}_{K}\mathbf{X}_{K}.\label{A_K}
\end{equation}
Since  $K<M$ must hold, this system is now an overdetermined system of linear equations. The reduced measurement matrix $\mathbf{A}_{K}$ would be formed with  the positions
of nonzero
samples $k\in \mathbb{K}$. It directly follows from matrix $\mathbf{A}$
when the columns corresponding to the zero-valued elements in
$\mathbf{X}$ are omitted. The reconstructed $\mathbf{X}_{K}$, with the assumed/known/determined nonzero positions, is a solution of the least square problem,
 \begin{equation}
 	\mathbf{X}_{K}=(\mathbf{A}_K^H\mathbf{A}_K)^{-1}\mathbf{A}^H_{K}\mathbf{y}.\label{A_K}
\end{equation}

The condition for this reconstruction is the invertibility of the matrix $\mathbf{A}_K^H\mathbf{A}_K$. This condition is much weaker than  the condition for a unique determination of the positions of nonzero elements in $\mathbf{X}$, $k\in \mathbb{K}$ that will be considered next.

\section{Unique Reconstruction} 

\textit{The $K$-sparse solution, whose elements are contained in the vector $\mathbf{X}_{K}$, is unique if all $\mathbf{A}_{2K}$ submatrices, corresponding to a $2K$-sparse signal  and
obtained from the measurement matrix $\mathbf{A}$, are such that all matrices $\mathbf{A}_{2K}^H\mathbf{A}_{2K}$ are invertible.} 

The contradiction will be used to prove this statement, being the basis for the derivation of the new limit for the sparsity.   Assume that two different $K$-sparse solutions exist for the vector $\mathbf{X}$.  Denote the nonzero elements of these solutions by $\mathbf{X}_K^{(1)}$
and $\mathbf{X}_K^{(2)}$. The nonzero elements in $\mathbf{X}_K^{(1)}$ correspond to the positions $k\in \mathbb{K}_1$ in the original vector $\mathbf{X}$, while  $\mathbf{X}_K^{(2)}$ contains the nonzero elements of  vector $\mathbf{X}$, positioned at $k\in \mathbb{K}_2$. Assume that both of these two solutions satisfy the measurements equation, that is,
$$\mathbf{A}_K^{(1)}\mathbf{X}_K^{(1)}=\mathbf{y} \text{ \   and \ } 
\mathbf{A}_K^{(2)}\mathbf{X}_K^{(2)}=\mathbf{y},$$ 
where $\mathbf{A}_K^{(1)}$
and $\mathbf{A}_K^{(2)}$ are submatrices of the measurement matrix $\mathbf{A}$
of size $M\times K$. They correspond to the nonzero elements in vectors $\mathbf{X}_K^{(1)}$
and $\mathbf{X}_K^{(2)}$, respectively. We can rewrite these two equations by adding zeros at the corresponding zero positions of other vectors, as
\begin{gather}
\begin{bmatrix}\mathbf{A}_K^{(1)} ~~~ \mathbf{A}_K^{(2)}
\end{bmatrix}\begin{bmatrix}\mathbf{X}_K^{(1)} \\ \mathbf{0}_K \end{bmatrix}
=\mathbf{y} \text{~~~and~~~}
\begin{bmatrix}\mathbf{A}_K^{(1)} ~~~ \mathbf{A}_K^{(2)}
\end{bmatrix}\begin{bmatrix} \mathbf{0}_K \\\mathbf{X}_K^{(2)}\end{bmatrix}
=\mathbf{y}.
\end{gather} 
If we subtract these two equations we get
\begin{gather}
\begin{bmatrix}\mathbf{A}_K^{(1)} ~~~ \mathbf{A}_K^{(2)}
\end{bmatrix}\begin{bmatrix}\mathbf{X}_K^{(1)} \\ -\mathbf{X}_K^{(2)} \end{bmatrix}
=\mathbf{0}. \label{SYS2K}
\end{gather} 
We arrived at the homogeneous system of equations. It is known that this system does not have nonzero solutions for the elements of $\mathbf{X}_K^{(1)}$ and $\mathbf{X}_K^{(2)}$
if the rank of matrix $\mathbf{A}_{2K}= \begin{bmatrix}\mathbf{A}_K^{(1)}
~~~ \mathbf{A}_K^{(2)}
\end{bmatrix}$ is equal to $2K$, meaning that $\mathbf{A}_{2K}^H\mathbf{A}_{2K}$ is invertible. If all possible submatrices $\mathbf{A}_{2K}$, for all possible combinations of nonzero element positions, of the measurement matrix $\mathbf{A}$
are such that $\mathbf{A}_{2K}^H\mathbf{A}_{2K}$ are invertible then two distinct solutions whose sparsity is $K$ cannot exist. This means that the
solution of the compressive sensing problem is unique.  Note that there are $\binom{N}{2K}$ submatrices $\mathbf{A}_{2K}$, and the combinatorial approach to this problem is not computationally feasible.

\section{Coherence}

The coherence
index of a matrix $\mathbf{A}$ is defined as the maximum absolute value
of
the normalized scalar product of its two columns, that is, \cite{CohD}
\[
\mu=\max\left\vert \mu_{mk}\right\vert \text{, for }m\neq k
\]
where the elements $\mu_{mk}$ are defined by
\begin{equation}
\mu_{mk}=\frac{1}
{||\mathbf{a}_m||_2||\mathbf{a}_k||_2} \sum_{i=0}^{M-1}a_{m}(i)a_{k}^{\ast}(i)=\frac{(\mathbf{a}_m,\mathbf{a}_k)}
{||\mathbf{a}_m||_2||\mathbf{a}_k||_2} 
\end{equation}
and $a_{k}(i)$ are the elements of the $i$th row and $k$th column, denoted by $\mathbf{a}_k$, of the measurement  matrix $\mathbf{A}$. If the measurement matrix is energy normalized, $||\mathbf{a}_k||^2_2=\sum_{i=0}^{M-1}\left\vert
        a_{k}(i)\right\vert ^{2}=1$, then 
\begin{equation}
\mu_{mk}=\sum_{i=0}^{M-1}a_{m}(i)a_{k}^{\ast}(i)=(\mathbf{a}_m,\mathbf{a}_k).
\end{equation}        
Notice that $\mu_{mk}$, are the elements of matrix $\mathbf{A}^H\mathbf{A}$. For $m \ne k$ we get the off-diagonal elements of the matrix $\mathbf{A}^H\mathbf{A}$, which is normalized so that  its diagonal elements assume unit value.

This coherence index plays
a crucial role in the measurement matrix design. The coherence
index should be as small as possible, meaning that the incoherence is
a desirable property for the measurement matrix \cite{INg}. With smaller values of the coherence index
the matrix defined by $\mathbf{A}^H\mathbf{A}$ has lower off-diagonal elements and is closer to the identity matrix. 

\section{Review of the Gershgorin Disk Theorem \cite{GSH}} The matrix $\mathbf{A}_{2K}^H\mathbf{A}_{2K}$ is invertible if its determinant is nonzero. This condition is equivalent to the condition that all eigenvalues of matrix $\mathbf{A}_{2K}^H\mathbf{A}_{2K}$, for all possible combinations of $2K$ nonzero element positions, are nonzero.  
The eigenvalue/eigenvector relation for a matrix $\mathbf{A}_{2K}^H\mathbf{A}_{2K}$ is defined by
$$(\mathbf{A}_{2K}^H\mathbf{A}_{2K})\mathbf{u}=\lambda\mathbf{u}$$
where $\mathbf{u}$ denotes the eigenvector corresponding to the eigenvalue $\lambda$. Since the eigenvector belongs to the kernel of $\mathbf{A}_{2K}^H\mathbf{A}_{2K}-\lambda \mathbf{I}$ we can always assume that its maximum coordinate is equal to $1$, that is $u_i=\max_j(u_j)=1$ and $|u_j| \le 1$ for $j \ne i$. 

For the columns $k \in \{k_1,k_2,...,k_{2K}\}$ selected from the matrix $\mathbf{A}$,  the elements of matrix $\mathbf{A}_{2K}^H\mathbf{A}_{2K}$ are denoted by 
\begin{equation}
\mu_{k_ik_j}=\sum_{m=0}^{M-1}a_{k_i}(m)a_{k_j}^{\ast}(m)=(\mathbf{a}_{k_i},\mathbf{a}_{k_j}),
\end{equation}  
for $i,j=1,2,\dots,2K$.      
Now, we can rewrite the eigenvalue relation as
$$\sum_j\mu_{k_ik_j}u_j=\lambda u_i =\lambda \text{ \ or \  }\sum_{j,j\ne i}\mu_{k_ik_j}u_j=\lambda -\mu_{k_ik_i}$$
From this relation we can conclude (Gershgorin Disc Theorem result) 
\begin{equation}|\lambda -\mu_{k_ik_i}| \le \sum_{j,j\ne i}|\mu_{k_ik_j}u_j|\le \sum_{j,j\ne i}|\mu_{k_ik_j}|,\label{GerDis}
	\end{equation}
where the property  $|u_j| \le 1$ for $j \ne i$ is used. Considering the eigenvalue $\lambda$ as a variable and $\mu_{k_ik_j}$ as constants, we conclude that the last inequality describes a disc area in the complex domain of $\lambda$, with the center at $\mu_{k_ik_i}$ and a radius $ \sum_{j,j\ne i}|\mu_{k_ik_j}|$. The disc described by the relation in (\ref{GerDis}) does not include the point $\lambda=0$ if the radius is smaller than the distance of the center from the origin, that is, if
\begin{equation}
 \mu_{k_ik_i}>\sum_{j,j\ne i}|\mu_{k_ik_j}|.\label{MIII}
 \end{equation}
Therefore, if the condition in (\ref{GerDis}) is met, the matrix $\mathbf{A}_{2K}^H\mathbf{A}_{2K}$ does not have a zero-valued eigenvalue, and it is therefore invertible. 

For a normalized matrix $\mathbf{A}_{2K}^H\mathbf{A}_{2K}$, we have $\mu_{k_ik_i}=1$.

We have already concluded that the solution for a $K$-sparse vector is unique if for all possible submatrices $\mathbf{A}_{2K}$ the matrices $\mathbf{A}_{2K}^H\mathbf{A}_{2K}$ are invertible. Note that the off-diagonal elements of $\mathbf{A}_{2K}^H\mathbf{A}_{2K}$ are a subset of the off-diagonal elements of the matrix $\mathbf{A}^H\mathbf{A}$. The same holds for the diagonal elements. It means that the coherence $\mu$ of matrix $\mathbf{A}$ will be always greater than or equal to the coherence of any submatrix $\mathbf{A}_{2K}$.

\textit{The invertibility condition for all matrices $\mathbf{A}_{2K}^H\mathbf{A}_{2K}$, and the unique solution for a $K$ sparse vector $\mathbf{X}$, is achieved if 
$ 1>(2K-1)\mu$
or 
\begin{equation} K<\frac{1}{2}(1+\frac{1}{\mu}).
	\label{Un1}
\end{equation}	
}
The proof of this classical coherence index-based uniqueness condition follows from (\ref{MIII})  
for the normalized matrix $\mathbf{A}^H\mathbf{A}$. The inequality  
\begin{equation}1=\mu_{k_ik_i}>\sum_{j=1,j\ne i}^{2K}|\mu_{k_ik_j}| \label{CohmiB}
	\end{equation}    
is satisfied if $1>(2K-1)\mu$ since $\sum_{k=1,k\ne m}^{2K}|\mu_{k_ik_j}|<(2K-1)\mu.$ 

\section{Improved Bound} 
The coherence index bound is, by definition, pessimistic since it takes the worst value $\mu$ for all $\mu_{mk}$ in (\ref{CohmiB}). Like in \cite{CB}, when the coherence index was analyzed, we may improve the coherence index-based bound in the Gershgorin disc theorem derivation using the sum of the $(2K-1)$ largest absolute values instead of using $(2K-1)$ times the largest absolute value  $\mu$, that is 
 \begin{equation}1>\max_i\{\sum_{j=1,j\ne i}^{2K}|\mu_{k_ik_j}|\}. \label{CohmiBBB}
 \end{equation}   
Since all possible combinations can appear in the worst case, in order to avoid combinatorial approach, we can use the largest values over the complete matrix $\mathbf{A}^H\mathbf{A}$. Denote the sorted values of the elements in the columns (or rows) of this matrix by 
$$s(m,p)=\mathrm{sort}_k\{|\mu(m,k)\},$$
such that $s(m,1)\ge s(m,2) \ge \dots \ge s(m,N)$ then  
we can write
\begin{equation}1>\max_m\{(2K-1)\frac{1}{2K-1}\sum_{p=1}^{2K-1}s(m,p) \} \label{CohmiBS}
\end{equation}   
or with $\beta_{\mathbf{A}}(2K-1)=\max_m\{\mathrm{mean}_{p=1,2\dots,2K-1}\{s(m,p)\}\}$,
\begin{equation}1>(2K-1)\beta_{\mathbf{A}}(2K-1) \label{CohmiBSS}
\end{equation}
Finally we get
\begin{equation}K<\frac{1}{2}(1+\frac{1}{\beta_{\mathbf{A}}(2K-1)}),\label{Un2}
	\end{equation}
where $\beta_{\mathbf{A}}(2K-1)$ is the mean value of $(2K-1)$ the largest  elements (in absolute value) of matrix $\mathbf{A}^H\mathbf{A}$ within one row/column. The implicit inequality is easily solved by checking for the sparsity values $K=1$, $K=2$, and so on, until the inequality in (\ref{Un2}) is still satisfied. 

Next, we will compare this bound with other derived bounds. It is obvious that this bound can improve the standard coherence index-based bound since $\mu\ge \beta_{\mathbf{A}}(2K-1)$, that is 
$$K<\frac{1}{2}(1+\frac{1}{\mu})\le \frac{1}{2}(1+\frac{1}{\beta_{\mathbf{A}}(2K-1)
}).$$

Next we can conclude that the bound in (\ref{Un2}) will be larger or equal to the one obtained in \cite{CB} using the average of the $(2K-1)$ largest values within the whole matrix $\mathbf{A}^H\mathbf{A}$,  
\begin{equation}K<\frac{1}{2}(1+\frac{1}{\alpha_{\mathbf{A}}})\le \frac{1}{2}(1+\frac{1}{\beta_{\mathbf{A}}(2K-1)}).\label{Un3}
\end{equation}
Finally, the bound derived here is compared with one  derived in \cite{CB}, when the maximum values in two rows are used, which is defined by  
\begin{gather}
	K < \frac{1+\beta_{\mathbf{A}}(K-1)}{\beta_{\mathbf{A}}(K-1)+\gamma_{\mathbf{A}}(K)}. \label{Un4}
\end{gather} 
 We cannot decisively conclude which one of the bounds in (\ref{Un2}) or (\ref{Un4}) is better since two different rows are used in the calculation of (\ref{Un4}). In the examples that will be presented next, the inequality (\ref{Un2}) produced higher sparsity bound than (\ref{Un4}) in all considered cases. 

All the previous bounds produce the same result for the equiangular tight frame (ETF) measurement matrices, when all $|\mu_{k_i k_j}|=\mu$ are equal for any $k_i\ne k_j$, and  
$\beta_{\mathbf{A}}(K-1)=\gamma_{\mathbf{A}}(K)=\alpha_{\mathbf{A}}=\mu$.

Finally, note that while the limit derived in \cite{CB} \textit{guarantees successful reconstruction} using the matching pursuit approach, the relaxed condition derived in this paper \textit{guarantees only the existence} of a unique solution.

\section{Numerical Examples} 
\begin{figure}
  	[ptb]
  	\begin{center}
  		
  (a)\includegraphics[scale=0.24]{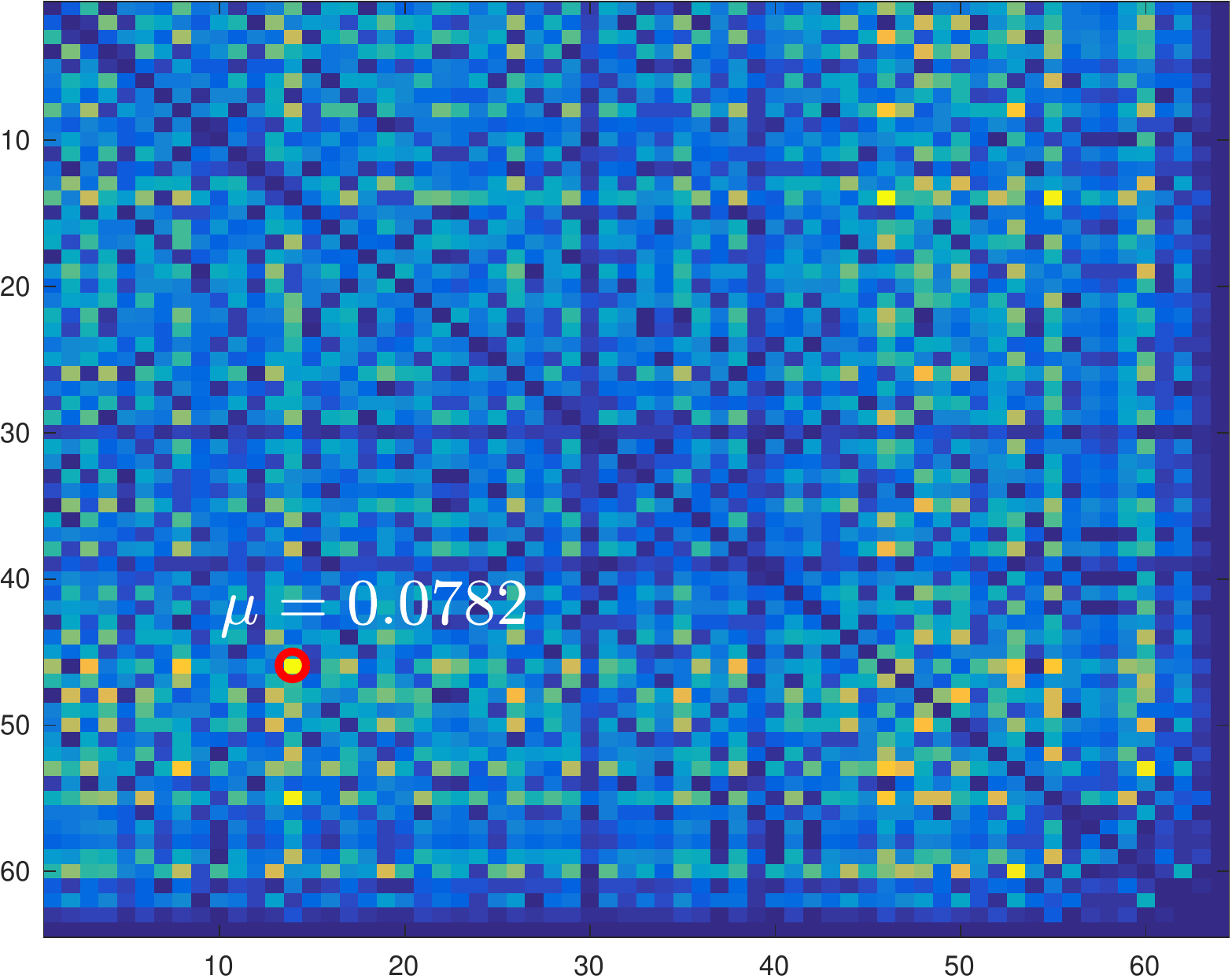}
   \includegraphics[scale=0.24]{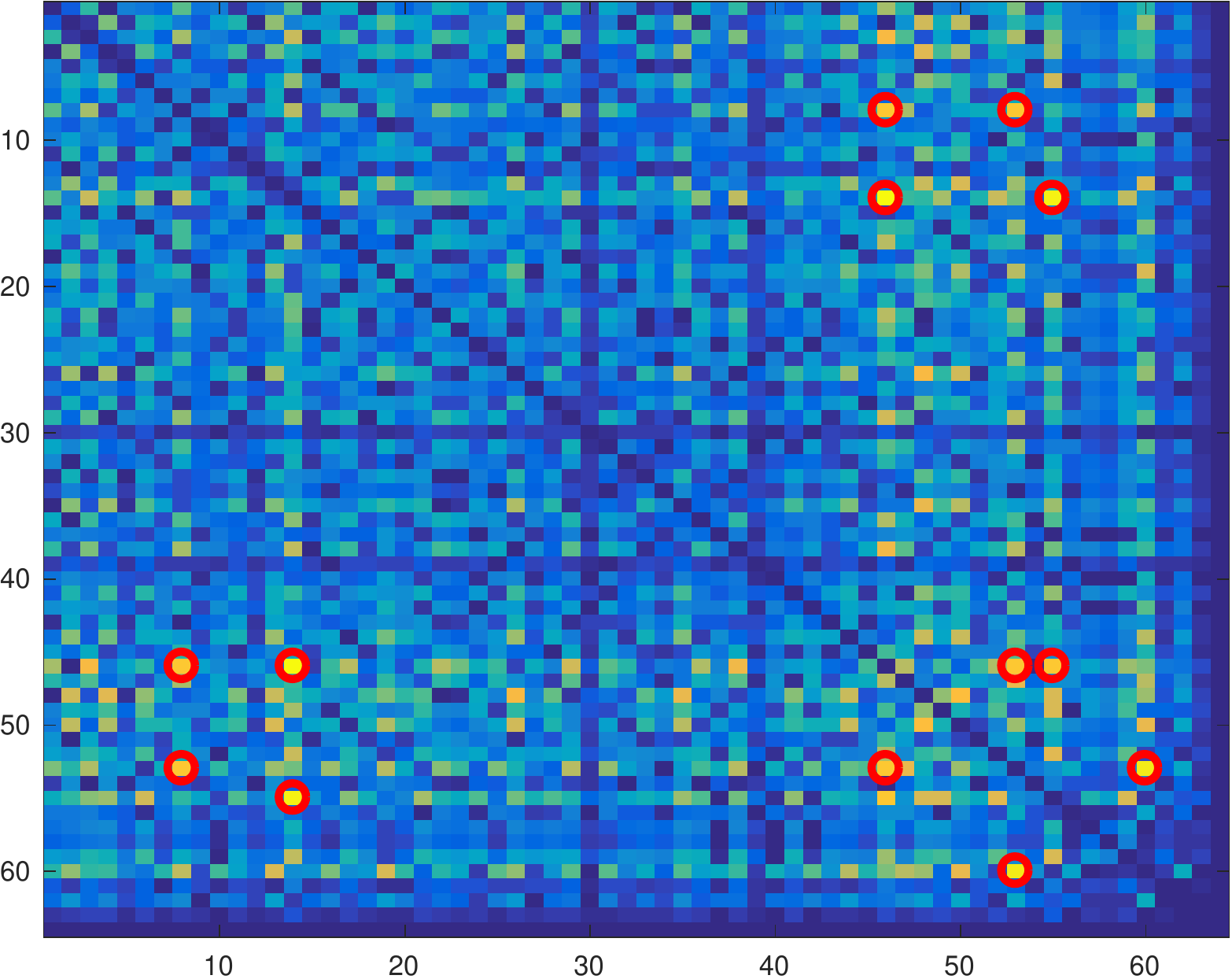}(b)
   
   (c)\includegraphics[scale=0.24]{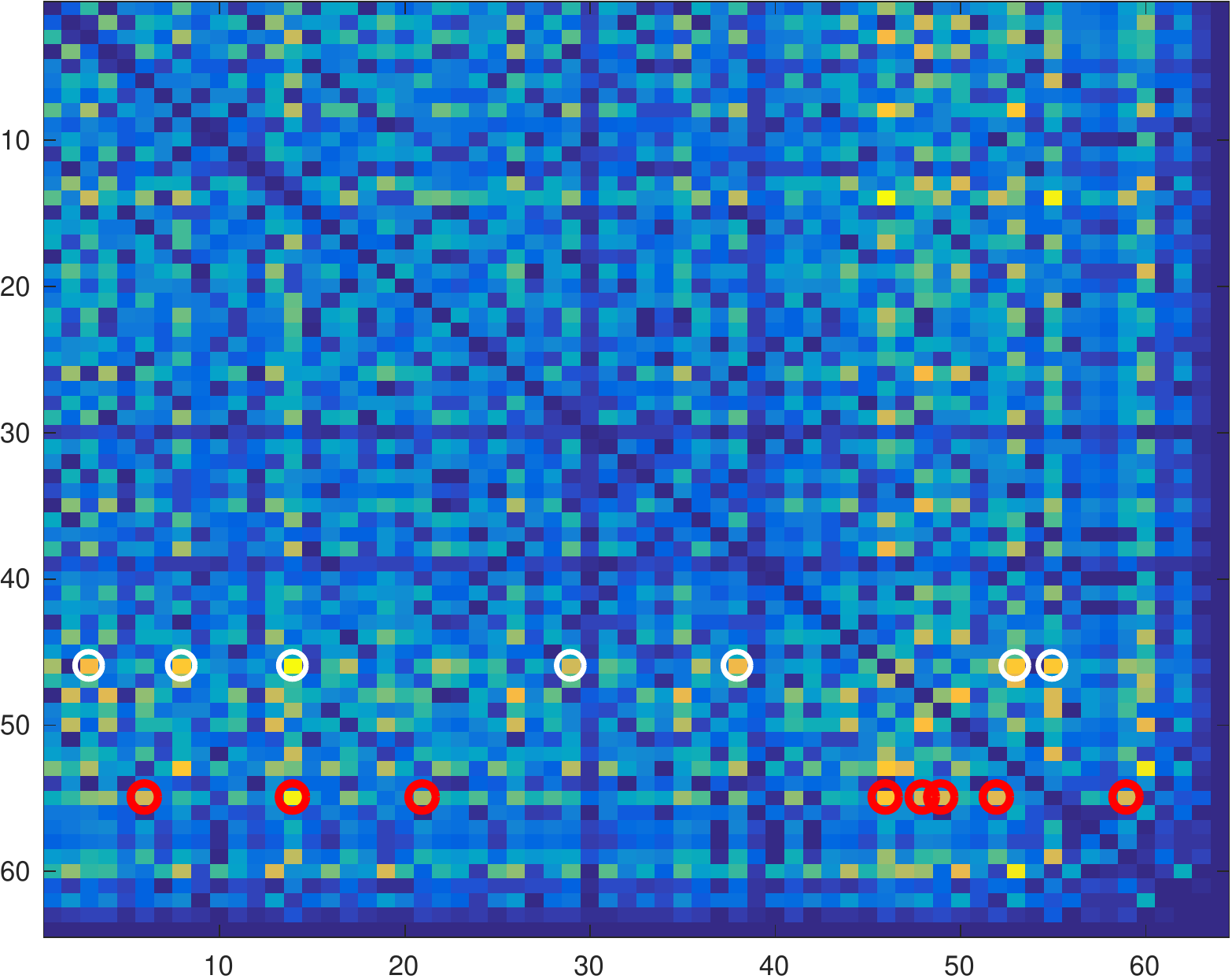} 
     \includegraphics[scale=0.24]{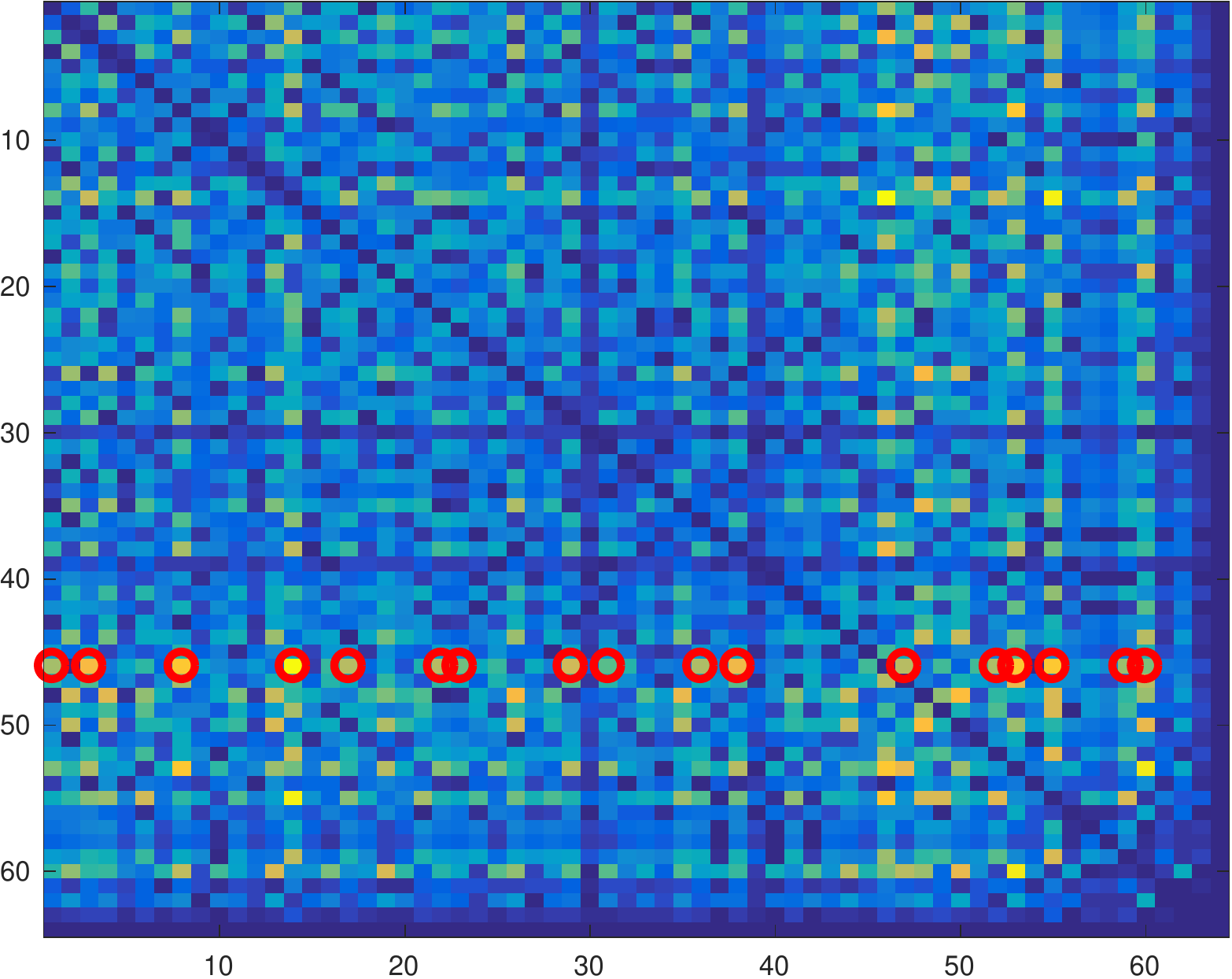}(d)
  		
  		\caption{The elements of matrix $\mathbf{A}^H\mathbf{A}$ used for the calculation of various bounds. (a) The coherence index, as the largest absolute off-diagonal element $\mathbf{A}^H\mathbf{A}$, used in the sparsity bound in (\ref{Un1}), marked with red circle.  (b) The largest absolute values of elements in $\mathbf{A}^H\mathbf{A}-\mathbf{I}$ used to calculate   $\alpha_{\mathbf{A}}$ and the bound in (\ref{Un3}). (c) The largest absolute values in $\mathbf{A}^H\mathbf{A}-\mathbf{I}$ used to calculate   $\beta_{\mathbf{A}}(K-1)$ (encircled using a white line) and $\gamma_{\mathbf{A}}(K)$ (encircled using a red line) used in the bound in (\ref{Un4}). (d) The largest absolute values in $\mathbf{A}^H\mathbf{A}-\mathbf{I}$ used to calculate   $\beta_{\mathbf{A}}(2K-1)$ and the bound in (\ref{Un2}). }\label{graph_SQ}
  	\end{center}
  \end{figure}

The limit for the sparsity was tested on several measurement matrices, including the partial graph Fourier transform (GFT) matrix , the partial DFT matrix, the partial DCT matrix, and a random Gaussian measurement matrix. 
\begin{itemize}
	\item For a partial DFT matrix $\mathbf{A}$ of dimension $124\times 128$ the sparsity limit obtained with the standard coherence index relation (\ref{Un1}) is $K<16.63$. For the limits (\ref{Un3}) and (\ref{Un4}) we get $K<16.63$ and $K<19.20$, respectively. For the limit in (\ref{Un2}) we get $K<23.54$. The proposed result improves the classical coherence index bound for almost $50\%$.  
	
	\item  For a Gaussian measurement matrix $\mathbf{A}$ of dimension $900\times 1000$ we get $K<16.63$ as the classical limit and $K<3.59$ and $K<4.48$, as the bounds in (\ref{Un3}) and (\ref{Un4})  respectively. With  (\ref{Un2}) we get $K<4.84$.
	
	\item For a partial DCT matrix of the size $124 \times 128$ we get $K<9.05$, $K<9.77$, $K<12.47$, and $K<15.11$, with the bounds defined by (\ref{Un1}), (\ref{Un3}), (\ref{Un4}), and (\ref{Un2}), respectively.   
	
	\item For a partial GFT matrix of a graph with $N=64$ vertices, given in \cite{CB}, the classical coherence index relation produces $K<6.89$. The bounds in (\ref{Un3}) and (\ref{Un4}) produce $K< 7.46$ and $K<8.31$, while the bound in (\ref{Un2}) produces $K<9.22$, as illustrated in Fig. \ref{graph_SQ}. 
\end{itemize}	
  
\section{Conclusion} An improved bound for the reconstruction limit has been recently proposed based on the coherence index analysis. In this paper, this bound is further relaxed by considering the existence of the unique solution only and using the Gershgorin disc theorem.   
%

\vskip5pt

\noindent L. Stankovic (\textit{Montenegrin Academy of Sciences and Arts (CANU), EE Department, University of Montenegro, Podgorica, Montenegro})
\vskip3pt

\noindent E-mail: ljubisa@ucg.ac.me

\end{document}